# Diaphragm as an anatomic surrogate for lung tumor motion

**Laura I. Cerviño, Alvin. K. Y. Chao, Ajay Sandhu, and Steve B. Jiang**

Department of Radiation Oncology, University of California San Diego, 3855 Health Sciences Dr, La Jolla, CA 92037-0843

E-mail: sbjiang@ucsd.edu

Lung tumor motion due to respiration poses a challenge in the application of modern three-dimensional conformal radiotherapy. Direct tracking of the lung tumor during radiation therapy is very difficult without implanted fiducial markers. Indirect tracking relies on the correlation of the tumor's motion and the surrogate's motion. The present paper presents an analysis of the correlation between the tumor motion and the diaphragm motion in order to evaluate the potential use of diaphragm as a surrogate for tumor motion. We have analyzed the correlation between diaphragm motion and superior-inferior lung tumor motion in 32 fluoroscopic image sequences from 10 lung cancer patients. A simple linear model and a more complex linear model that accounts for phase delays between the two motions have been used. Results show that the diaphragm is a good surrogate for tumor motion prediction for most patients, resulting in an average correlation factor of 0.94 and 0.98 with each model respectively. The model that accounts for delays leads to an average localization prediction error of 0.8mm and an error at the 95% confidence level of 2.1mm. However, for one patient studied, the correlation is much weaker compared to other patients. This indicates that, before using diaphragm for lung tumor prediction, the correlation should be examined on a patient-by-patient basis.

## 1. Introduction

It has been shown that modern three-dimensional radiation therapy is more successful at treating early stage lung cancer than older two-dimensional radiation therapy (Fang *et al.*, 2006). There is, however, a concern related to respiratory tumor motion. It has been shown that the motion amplitude can be clinically significant (~2-3 cm) (Keall *et al.*, 2006), depending on tumor location and individual patients. Tumor motion may greatly degrade the effectiveness of conformal radiotherapy for the management of lung cancer, especially when the treatment is done in a hypofractionated or single-fraction manner.

Various tumor motion compensation strategies have been investigated (Keall *et al.*, 2006; Jiang, 2006a). Advanced motion compensation techniques, such as beam gating (Jiang, 2006b) or beam tracking (Keall *et al.*, 2001), rely on the precise knowledge of tumor position. There are mainly four different methods to determine tumor location during the treatment: 1) direct fluoroscopic tumor tracking (Cui *et al.*, 2007; Xu *et al.*, 2007; Xu *et al.*, 2008); 2) fluoroscopic tracking of the implanted fiducial markers (Sharp *et al.*, 2004; Tang *et al.*, 2007); 3) inference of the tumor position from anatomic surrogates such as abdomen surface (Berbeco *et al.*, 2005b; Berbeco *et al.*, 2006; Wu *et al.*, 2008); and 4) non-radiographical tumor tracking through implanted transponders (Seiler *et al.*, 2000; Balter *et al.*, 2005).

Tumor tracking techniques based on implanted fiducial markers or wireless transponders, when applied to lung cancer, suffer from various clinical problems, such as the risks of pneumothorax (Geraghty *et al.*, 2003; Arslan *et al.*, 2002) and marker migration (Nelson *et al.*, 2007). Therefore, these techniques are not suitable for lung tumor tracking. Techniques for direct fluoroscopic lung tumor tracking are still under development and there is a long way to go before its clinical use (Cui *et al.*, 2007; Xu *et al.*, 2007; Xu *et al.*, 2008; Lin *et al.*, 2009). Tumor localization based on anatomic surrogates has been implemented clinically for gated lung cancer radiotherapy, using commercial products such as the Real-Time Position Management (RPM) system (Varian Medical Systems, Inc, Pala Alto, CA, USA) (Jiang, 2006b). This kind of techniques is accurate only when there is a good correlation between lung tumor motion and the motion of the anatomic surrogate. Therefore, it is crucial to have a clear understanding on the correlation between tumor and a particular surrogate before this surrogate is used to derive the tumor location.

The most commonly used anatomic surrogate is the antero-posterior (AP) motion of the abdominal surface (Jiang, 2006b; Kanoulas *et al.*, 2007; Wu *et al.*, 2008; Ruan *et al.*, 2008). Vedam *et al* have shown that the abdominal surface motion correlates strongly with diaphragm motion (Vedam *et al.*, 2003). However, some other studies show that the correlation between tumor motion and abdominal surface motion or between abdominal surface motion and diaphragm motion depends on individual patients and treatment fractions and cannot be generalized (Bruce, 1996; Hoisak *et al.*, 2004, 2006; Tsunashima *et al.*, 2004). When abdominal surface motion is used to generate gating signal, the tumor residual motion can be large (Berbeco *et al.*, 2005b; Berbeco *et al.*, 2006) and about 30% of the time the radiation beam will miss the target (Wu *et al.*, 2008). Patient breath coaching or updating the internal/external correlation using low frequency x-ray imaging can improve the accuracy gated radiotherapy based on abdominal surface motion (Jiang, 2006b; Kanoulas *et al.*, 2007; Wu *et al.*, 2008).

Intuitively, anatomic surrogates inside the thorax, such as carina or diaphragm, should correlate better with lung tumor motion compared to abdominal surface. Van der Weide *et al.* found that there is a good correlation between carina and lung tumor motion and concluded that carina is a better internal surrogate than diaphragm (van der Weide *et al.*, 2008). However, when estimating the correlation between diaphragm and lung tumor motion, they did not take in account the phase shift between two motions, which underestimates the correlation. Zhang *et al.* evaluated tumor position prediction from diaphragm with a linear model that compensated for phase shift between tumor and diaphragm motions (Zhang *et al.*, 2007). They performed the study only for 4 patients. Both studies are based on 4D-CT data, which consists on only one fictitious breathing cycle. We believe that the diaphragm is more suitable than carina for clinical application due to its high visibility in fluoroscopic images, if there is a good correlation between diaphragm and lung tumor motion. Therefore, we feel there is a need to carefully and comprehensively examine the correlation of lung tumor position with diaphragm position using image sequences consisting on many breathing cycles.

The goal of this project is to analyze the correlation between the motions of the diaphragm and the lung tumor in order to evaluate whether diaphragm can be used as an internal anatomic surrogate for predicting lung tumor position. Diaphragm, as opposed to tumor, can be easily detected in fluoroscopic images (Chen *et al.*, 2001; Berbeco *et al.*, 2005a; Vedam *et al.*, 2003).

**2. Methods and Materials**

*2.1 Patient image data*

A retrospective analysis was performed based on 32 AP fluoroscopic sequences from 10 different lung cancer patients who underwent hypofractionated radiotherapy at the Moores Cancer Center, University of California San Diego (UCSD). The fluoroscopic images were acquired before or after treatment, at different treatment fractions, using an on-board x-ray imaging (OBI) system (Varian Medical Systems, Inc., Palo Alto, CA, USA). Each fluoroscopic sequence consists of an average of 513 images (320 to 701) that were acquired at a rate of 15 frames per second. Each image has 1024x768 pixels, with pixel size of either 0.243mm or 0.259mm, depending on the sequence. Tumor position was manually marked in each frame by an expert observer, and was considered as the ground truth. When tumor was not easily identified in the fluoroscopic sequence, an anatomical feature close to the tumor and easy to detect was marked instead. The position of diaphragm was automatically detected by thresholding the image and applying a maximum gradient algorithm. The position of the apex of the diaphragm was recorded for all the sequences and frames.

*2.2 Overview of the model construction*

In order to observe the correlation between diaphragm and tumor motions, two different models for tumor motion prediction based on diaphragm motion were developed. First we used a linear model where tumor position was derived from the concurrent diaphragm position (Model 1). Then, to take into account the possible phase shift between tumor motion and diaphragm motion, we developed a more complex linear model where the concurrent diaphragm position and an optimal set of past positions of the diaphragm were considered as surrogates (Model 2). Two different analyses were performed on both models: 1) a goodness of fit study, where the models were built based on all the images of the sequence, i.e., training set consisted on all the images in the sequence; and 2) a prediction power study, where the models were built based on the first 200 images (training set), corresponding to 13.3 seconds, and prediction errors were computed during the rest of the sequence (testing set). In order to eliminate the noise inherent to the manual identification of the tumor position, tumor trajectories were filtered (in time) using a 5-point median filter. All our algorithms have been implemented on Matlab 7.4 platform.

*2.3 Model construction*

Two different sets of surrogates have been considered for each patient, resulting in two different models (Model 1 and Model 2). Both models follow a procedure similar to that described by Zhang *et al* (2007). The objective of these models is to establish the linear relationship between tumor displacement $\tilde{y}(t)$ and surrogate displacement $\tilde{\mathbf{s}}(t)$:

$$\tilde{y}(t) = \mathbf{B}\tilde{\mathbf{s}}(t), \tag{1}$$

where $\tilde{\mathbf{s}}(t) = [\tilde{s}_1(t), \tilde{s}_2(t), ..., \tilde{s}_N(t)]^T$ is the vector with the surrogate displacements (in this section of the paper vectors and matrices are indicated in bolded font), *t* denotes time, and *N* is the number of surrogates used in the model. $\tilde{y}(t)$ and $\tilde{\mathbf{s}}(t)$ are displacements of the tumor position $y(t)$ and surrogates position $\mathbf{s}(t)$ with respect to their mean position $\bar{y}$ and $\bar{\mathbf{s}}$ (that is, $\tilde{\mathbf{s}}(t) = \mathbf{s}(t) - \bar{\mathbf{s}}$ and $\tilde{y}(t) = y(t) - \bar{y}$). Note that, while Zhang *et al* (2007) perform a non-rigid registration of the entire images in order to find the deformation field for all the pixels in the image, we have only considered the motions of tumor centroid and surrogates. Still the same method has been used here because of its generalizability to more complex structures, such tracking different tumors or different points within the image at the same time. Constructing a vector **p** that contains the tumor superior-inferior (SI) position *y* and the surrogates $\mathbf{p}(t)=[y(t), s_1(t), s_2(t), ... , s_N(t)]^T$, and performing a Principal Component Analysis (PCA) on $\mathbf{p}(t)$ (notice that there are only as many as *N* principal components), we can express any $\mathbf{p}(t)$ as a linear combination of the eigenvectors obtained from the PCA:

$$\mathbf{p}(t) = \bar{\mathbf{p}} + \sum_{k=1}^{N} w_k(t)\mathbf{e}_k, \tag{2}$$

where $\mathbf{e}_k$ are the eigenvectors of the analysis (components), $w_k$ are the corresponding coefficients in the linear combination, and $\bar{\mathbf{p}} = [\bar{y}, \bar{s}_1, ..., \bar{s}_N]$ is the mean value of $\mathbf{p}(t)$. Eq. (2) can be written in matrix form as:

$$\tilde{\mathbf{p}} = \mathbf{p}(t) - \bar{\mathbf{p}} = \mathbf{EW} \tag{3}$$

where $\mathbf{E} = [\mathbf{e}_1, ..., \mathbf{e}_N]$ and $\mathbf{W} = [\mathbf{w}_1, ..., \mathbf{w}_N]$. Noting that $\tilde{\mathbf{p}} = [\tilde{y}, \tilde{s}_1, ..., \tilde{s}_N]^T$, we can split Eq. (3) into two different equations:

$$\tilde{\mathbf{y}} = \mathbf{E}_y \mathbf{W}$$
$$\tilde{\mathbf{s}} = \mathbf{E}_s \mathbf{W} \qquad (4)$$

where $\mathbf{E}_y$ and $\mathbf{E}_s$ are the upper row and the $N$ lower rows of $\mathbf{E}$ respectively. Eliminating $\mathbf{W}$ in Eq. 4 and assuming that $\mathbf{E}_s$ is invertible, we obtain:

$$\tilde{y} = \mathbf{E}_y \mathbf{E}_s^{-1} \tilde{\mathbf{s}}(t) = \mathbf{B}\tilde{\mathbf{s}}(t) \quad , \qquad (5)$$

which is the relationship we were looking for in Eq. (1).

**Model 1 (one surrogate)**: for the construction of the first model the position of the tumor $y_0$ at time $t_0$ is derived from only one surrogate, i.e., the position of the diaphragm $s_0$ at the same time instant (see Figure 1). The equation of the model is:

$$y_0 = a + bs_0, \qquad (5)$$

where the coefficients $a$ and $b$ are considered constant, and are obtained from the procedure indicated previously (note that $a = \bar{y}$ and $b = \mathbf{B}$). $a$ and $b$ are considered as constants in the model.

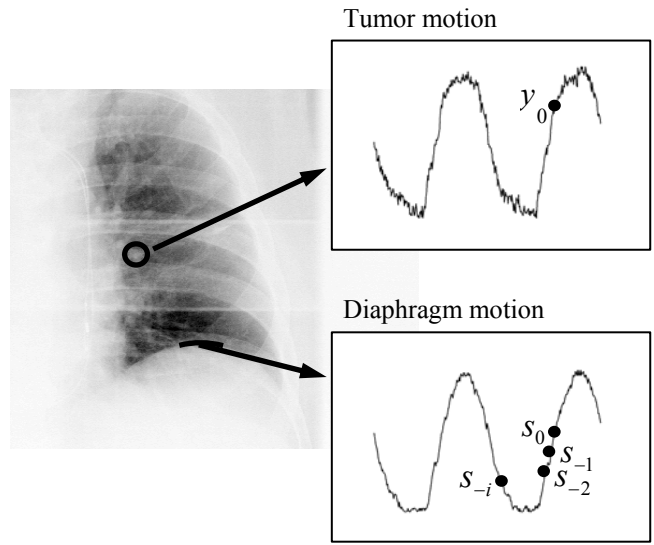

**Figure 1:** Representation of tumor and diaphragm motion variables used in the construction of models 1 and 2.

**Model 2 (optimal set of surrogates)**: for the construction of the second model, the position of the tumor at time $t_0$ is expressed as a linear function of an optimal set of surrogates that includes the current position of the diaphragm $s_0$ and an optimal combination of past diaphragm positions (Figure 1). The relationship between tumor position and diaphragm are expressed by:

$$y_0 = a + \sum_i^N b_i s_i, \qquad (6)$$

where the summation includes only the optimal set plus the current position. Again, coefficients *a* and $b_i$ are considered constant during the whole sequence ( $a = \bar{y}$ and $[b_1,...,b_N] = \mathbf{B}$ ). Adding the diaphragm position at past times to the model we account for possible phase shifts and shape differences between diaphragm and tumor motions.

We have selected a maximum N of 20, which allows for more than 1 second delay between motions. An exhaustive search among all the possible subsets of these surrogates to find the optimal subset would lead to about 0.5 million searches. Instead, a subset of surrogates close to optimal was obtained by the Sequential Forward Floating Search (SFFS) (Pudil *et al.*, 1994). Sequential feature selection algorithms search for features in a sequential deterministic manner to find a suboptimal subset of features with respect to an evaluation criterion *J*. Forward methods start with an empty subset and sequentially add new features to achieve optimality. SFFS is a feature selection technique that provides close to optimal solution at an affordable computational cost (Jain *et al.*, 2000). The SFFS method comprises the following steps:

   Step 1: Inclusion (Sequential Forward Search, SFS)
   Step 2: Conditional exclusion
   Step 3: Continuation of conditional exclusion

At Step 1 one feature among the available features is added to the surrogate subset, so that the new subset maximizes the evaluation criterion *J*. At Step 2 a conditional exclusion is performed, that is, the worst feature in the updated subset is excluded from the set if by doing so a best evaluation criterion *J* is achieved. At Step 3 conditional exclusion steps continue until no further improvement in *J* is achieved. These 3 steps are iteratively repeated until no further improvement in *J* is found.

In this study, the selection criterion *J* used for the feature selection in the SFFS method is the correlation coefficient obtained between the tumor position trajectory and the predicted position given by the selected surrogates during the training set.

*2.4 Goodness of fit*

In order to evaluate correlation, a Pearson's correlation test was performed. Models 1 and 2 were built based on the diaphragm position and the tumor position in the *whole* fluoroscopic sequence, for each sequence individually. The correlation between the real tumor position and the predicted tumor position given by each models was calculated, as well as the mean error $\bar{e}$ and the maximum error at the 95% confidence interval $e_{95}$ of the prediction.

In addition, Model 1 has been used for comparison of model parameters between different sequences of the same patient. We focus on variability of the slope *b* in Eq. (5)

of these models and the correlation coefficient of tumor and diaphragm motions for each patient. Variability has been computed as a percentage by calculating the standard deviation of the parameters divided by the average, and multiplying by 100.

*2.5 Prediction power*

In addition to calculating the correlation coefficient, an evaluation of the predictability of the models was performed. In this case, the first 200 frames of each sequence were used as the training set to build Models 1 and 2, and the remaining frames in the same session were used as testing data. For every sequence, models built with the training set were applied to predict tumor position in the testing set. Mean prediction error $\bar{e}$ and 95-percentile prediction error $e_{95}$ are computed in the testing set in each sequence.

**3. Results**

*3.1 Data analysis*

Table 1 shows, for each patient, the number of sequences analyzed, the average tumor motion range, the average delay between diaphragm and tumor motion, and the average distance between tumor and diaphragm. It can be observed that the average tumor motion ranged from 12 to 24mm, although individual sequence values ranged from 9mm to 35mm. The average motion range among all the patients and sequences was found to be 17mm. Delay between diaphragm and tumor motions was calculated with cross correlation of both signals (tumor and diaphragm trajectories). Mean delay and standard deviation are given for patients with more than one fluoroscopic sequence. Patients with several fluoroscopic sequences show a constant delay throughout all the series. Patient 2 shows a standard deviation of 0.15 sec, mainly due to the fact that in some sessions the patient remains without breathing for prolonged periods of time, which affects the calculation of the delay.

**Table 1:** Summary of patient data with number of sequences acquired for each patient, average tumor motion, and delay between diaphragm and tumor motion.

| Patient | No. of sequences | Average tumor motion (mm) | Delay (sec) | Average distance tumor-diaphragm (mm) |
|---|---|---|---|---|
| 1 | 5 | 12 | 0.17±0.04 | 46 |
| 2 | 6 | 13 | 0.34±0.15 | 41 |
| 3 | 3 | 23 | 0.40±0.0 | 20 |
| 4 | 2 | 17 | 0.00±0.0 | 81 |
| 5 | 1 | 24 | 0.27 | 3 |
| 6 | 3 | 16 | 0.00±0.0 | 2 |
| 7 | 9 | 20 | 0.00±0.0 | 48 |
| 8 | 1 | 13 | 0.00 | 14 |
| 9 | 1 | 18 | 0.13 | 9 |
| 10 | 1 | 12 | 0.07 | 3 |

*3.2 Correlation between diaphragm and tumor motions*

Figure 2 shows the general trend of model performance for most patients: Model 1 (red hollow circles) proves a good correlation between diaphragm and tumor motion, which can be further improved by adding surrogates from the diaphragm history in Model 2 (blue solid circles). A detail of the true target position and modeled position is shown. It can be seen that Model 2 corrects for a slight phase delay between diaphragm and target motions. The usual number of surrogates in the optimal subset ranges from 2 to 4 surrogates. Although different sequences of the same patient may use a different subset of surrogates, they are similar.

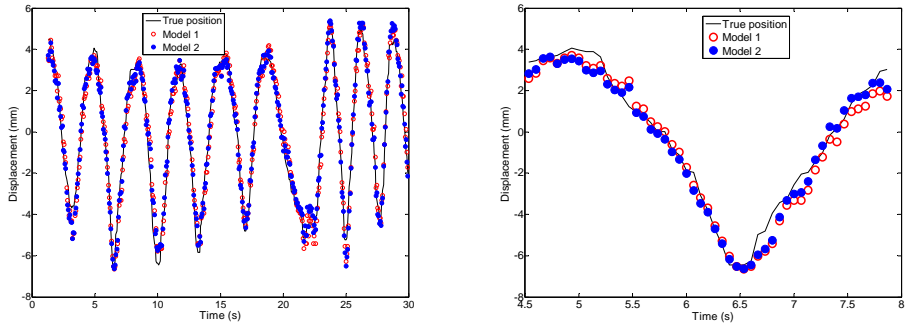

**Figure 2:** Real tumor position (solid line) and tumor position predicted by Model 1 (red hollow circles) and Model 2 (blue solid circles) for Patient 1. Left: entire fluoroscopic sequence. Right: detail of a subinterval of the sequence to observe the difference between the two models. Patient shows a good correlation between diaphragm and tumor motion. Model 2 corrects for a slight phase delay between tumor and diaphragm motions.

Table 2 shows the results of the study of the correlation between diaphragm position and tumor position given by each model (Model 1 and Model 2). The training set in this study is the whole sequence. Correlation coefficient obtained with each model, average error $\bar{e}$, and localization error at 95% confidence level $e_{95}$ are shown for each patient, as well as the average of all the sequences and patients. The average correlation coefficient between the tumor position and the estimated position given by Model 1 is 0.94, although it is always larger than 0.94 except for Patient3, who presents a correlation coefficient of 0.76. It can be observed that Model 2 gives an average correlation value of 0.98 (range 0.95-1.00). Model 2 gives smaller mean error ($p<0.01$) and 95-percentile error ($p<0.01$) and higher correlation coefficients ($p<0.03$) compared to Model 1. Model 2 gives an $e_{95}$ of less than 2.3mm except for Patient3.

**Table 2**: Correlation results for Model 1 and Model 2: average correlation factor $\rho$, average mean error $\bar{e}$, and average 95-percentile error $e_{95}$ are shown.

|         | Model 1 |        |           | Model 2 |        |           |
|---------|---------|--------|-----------|---------|--------|-----------|
| Patient | $\rho$  | $\bar{e}$ (mm) | $e_{95}$ (mm) | $\rho$ | $\bar{e}$ (mm) | $e_{95}$ (mm) |
| 1       | 0.95    | 0.9    | 1.9       | 0.99    | 0.4    | 1.0       |
| 2       | 0.97    | 0.7    | 1.8       | 0.98    | 0.6    | 1.4       |
| 3       | 0.76    | 3.6    | 8.0       | 0.95    | 1.4    | 3.3       |
| 4       | 0.96    | 0.8    | 2.0       | 0.97    | 0.7    | 1.6       |
| 5       | 0.95    | 1.7    | 4.1       | 0.98    | 1.0    | 2.3       |
| 6       | 0.99    | 0.6    | 1.3       | 1.00    | 0.3    | 0.8       |
| 7       | 0.98    | 0.7    | 1.7       | 0.99    | 0.6    | 1.5       |
| 8       | 0.94    | 0.7    | 1.8       | 0.95    | 0.7    | 1.7       |
| 9       | 0.98    | 1.0    | 2.1       | 0.99    | 0.7    | 1.8       |
| 10      | 0.95    | 0.9    | 2.2       | 0.97    | 0.8    | 1.7       |
| Average | 0.94    | 1.1    | 2.7       | 0.98    | 0.7    | 1.7       |

In patients with several series, it has been observed that parameters in Model 1 do not change greatly among fractions. The variability of the linear slope for Patients 1, 2, 4, 6 and 7 is 9.0%, 4.4%, 8.3%, 3.5% and 18% respectively. Variability in Patient 7 is higher because of one series where patient showed much deeper inspirations and much more irregular breathing than in all the other series. The average variability is 8.6%. Variability of the correlation coefficient is 1.1%, 1.8%, 3.0%, 2.9% and 2.6% respectively

*3.3 Prediction power*

To examine the prediction power of models for tumor locations based on diaphragm positions as surrogates, the first 200 frames of each sequence were used to build both models (training set) in each sequence independently. These models were used to predict tumor position in the remaining frames of the same sequence (testing set). The average correlation between real and predicted tumor position, and the average prediction errors $\bar{e}$ and $e_{95}$ of all sequences in each patient and the total average for all patients are shown in Table 3. It can be observed that correlation coefficients of the prediction are very similar to the ones previously obtained. On average, Model 2 improves tumor motion prediction $e_{95}$ over Model 1 by almost 1mm. The average prediction $e_{95}$ for all the sequences and all the patients is 2.1mm. The patient $e_{95}$ averages range form 1.1mm to 4.6mm, this last value corresponding to Patient3. If Patient3 were excluded from the study, the $e_{95}$ range would be 1.1-2.9mm.

**Table 3:** Average prediction error for each patient given by the each model.

|         | Model 1 |        |           | Model 2 |        |           |
|---------|---------|--------|-----------|---------|--------|-----------|
| Patient | $\rho$  | $\bar{e}$ (mm) | $e_{95}$ (mm) | $\rho$ | $\bar{e}$ (mm) | $e_{95}$ (mm) |
| 1       | 0.94    | 1.0    | 2.1       | 0.98    | 0.6    | 1.3       |
| 2       | 0.98    | 0.9    | 2.4       | 0.99    | 0.7    | 1.9       |
| 3       | 0.77    | 3.7    | 8.1       | 0.96    | 1.5    | 4.6       |
| 4       | 0.97    | 1.2    | 2.6       | 0.96    | 0.8    | 2.0       |
| 5       | 0.94    | 1.8    | 4.2       | 0.98    | 1.1    | 2.9       |
| 6 | 0.99 | 0.6 | 1.3 | 0.99 | 0.5 | 1.1 |
| 7 | 0.98 | 0.8 | 1.9 | 0.98 | 0.7 | 1.8 |
| 8 | 0.92 | 0.7 | 1.6 | 0.92 | 0.7 | 1.6 |
| 9 | 0.98 | 1.0 | 2.1 | 0.98 | 0.8 | 1.8 |
| 10 | 0.96 | 1.0 | 2.1 | 0.97 | 0.8 | 1.7 |
| Total Average | 0.94 | 1.3 | 2.8 | 0.97 | 0.8 | 2.1 |

Figure 3 shows the prediction error for models 1 and 2 for all the patients and all the sequences. Model 2 gives smaller errors than Model 1 in all the studied cases.

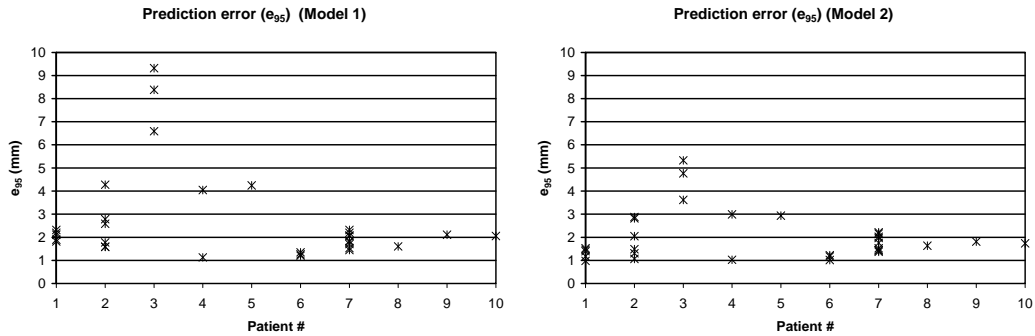

**Figure 3:** Prediction error for Model 1 (left) and Model 2 (right) in all the sequences

We have studied whether there is any correlation between motion range or tumor-diaphragm distance and prediction error. Figure 4 shows the 95-percentile error $e_{95}$ versus the motion range for all the fluoroscopic sequences, as well as a linear fit to this data. An increase of error with motion range is observed, although the R-square value of this fit is small. Figure 5 shows $e_{95}$ versus the distance between diaphragm and tumor mean positions for each sequence, and the linear fit to this data. The small R-square value of this fit indicates that there is no correlation between both magnitudes.

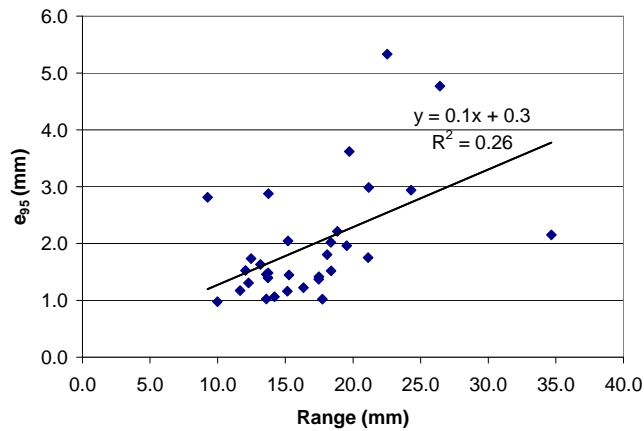

**Figure 4**: 95-percentile prediction error versus motion range for all the fluoroscopic sequences of all the patients using Model 2.

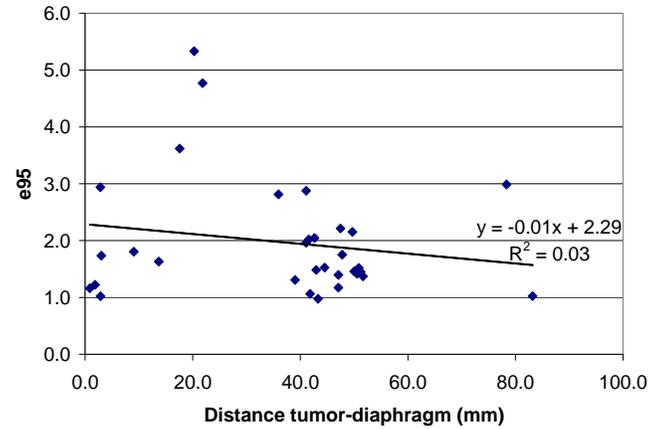

**Figure 5**: 95-percentile prediction error versus tumor-diaphragm distance for all the fluoroscopic sequences of all the patients using Model 2.

*3.4 Outlier Patient*

During the model evaluation and correlation study we observed that all the patients showed good correlation factors and good model predictions, except for Patient 3. After analyzing one of Patient3's sequences visually, we observed that the motion inside this patient's lung seems at times inconsistent. For example, at times some structures inside the lung move while the diaphragm remains static (two different snapshots of one of the sequences of Patient3 are shown in Figure 6). These two temporary uncorrelated motions, possibly due to coughing, led to low overall correlation factors and poor model predictions. Analysis of the two other sequences of the same patient did not lead to any observation of such inconsistency in motion. Results, however, although better than in the sequence previously mentioned, were still worse than for other patients (prediction $e_{95}$ with Model 2 was 5.3mm and 3.6mm). Figure 7 shows the motion signal of the tumor and the diaphragm in a sequence of Patient3. It can be observed that not only the amplitude and phase are different, but also the shape of both motions is different, with the tumor showing more pronounced peaks. Tumor trajectory cannot be represented with a simple linear model such as Model 1. Model 2 corrects for phase delay and shape, although still regions in the breathing cycle are not well represented. A non-linear model should be explored for this patient.

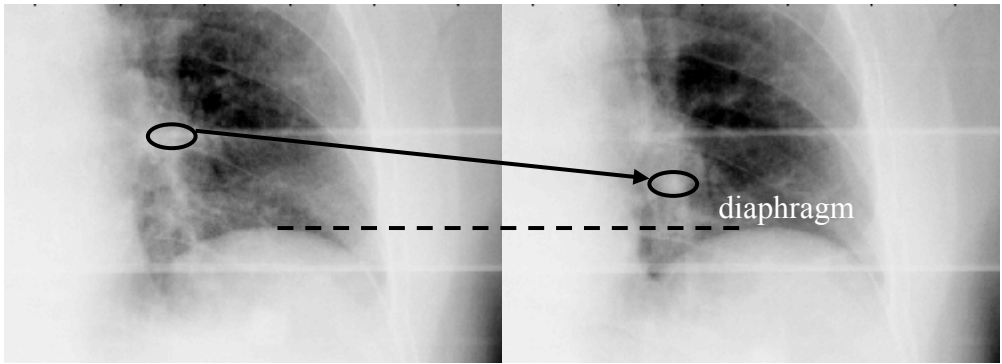

**Figure 6**: Two snapshots of a fluoroscopic sequence of Patient3 highlighting the change of some of the underlying structures while the diaphragm does not move.

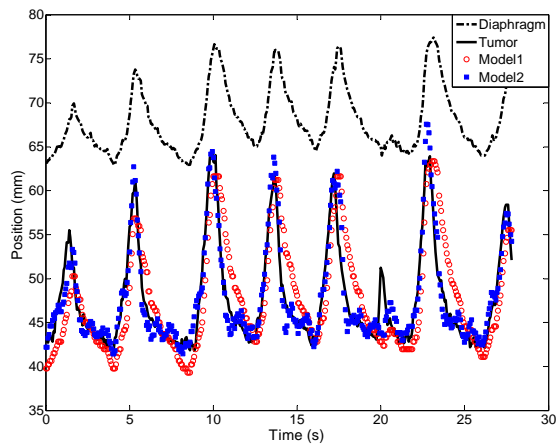

**Figure 7:** Diaphragm and real and estimated tumor positions for a sequence of Patient3.

**4. Discussion and conclusions**

This paper presents a study of the correlation between diaphragm motion and tumor craniocaudal motion based on 32 fluoroscopic sequences of 10 different lung cancer patients. The correlation factor has been shown to be very good, especially with Model 2 (which compensates for phase shifts), being always superior to 0.95, and leading to an average prediction error of 0.8mm and error at a 95% confidence level of 2.1mm. The small prediction error and the high correlation between diaphragm position and tumor position with the linear models used in this work lead to the conclusion that diaphragm motion is a good surrogate for indirect tumor position estimation. It should be noted that if non-linear models were considered in addition to linear models, results should improve, or at least not get worse. Gated treatment or tumor tracking can clearly benefit from the good correlation here found. While tumor is not always visible in fluoroscopic images and it cannot be directly tracked, the diaphragm has a high contrast edge that can be automatically identified at the required rate for a successful online estimation and treatment.

It has been observed that the correlation between diaphragm motion and tumor motion varies from patient to patient. Therefore a specific training set to build the model should be used for each individual. While performance of the models here developed show that in general there is a linear (and therefore predictable) relationship between diaphragm and tumor motion, this has not been the case for Patient3. Although adding new surrogates to the model seems to improve the results, error is still larger than desired. Therefore the appropriateness of a linear relationship between diaphragm and tumor motions should be assessed for every patient. Patients like Patient3 in this study, who present a difference in the shape of diaphragm motion trajectory and tumor motion trajectory that cannot be explained by a simple amplitude change and time delay, might benefit from the development of a non-linear model.

We would like to acknowledge that one limitation of the presented study is that it only considers craniocaudal tumor motion. Lateral and AP motions usually have smaller amplitude than superior-inferior motion, and lead to large noise in the marked trajectory when fluoroscopic images are manually marked. Although it is expected that the correlation of diaphragm motion and lateral and AP tumor motion is also good, a similar study where the 2D or 3D tumor position is known should be performed in the future.

In our prediction power study 200 images were used to build the prediction model. In contrast, other correlation studies normally use 4D-CT images, which consist on about 10 breathing phases (van der Weide *et al.*, 2008; Zhang *et al.*, 2007). One of the problems of these 4D-CT studies is that the scans at the diaphragm axial position and the scans at each tumor position correspond to different breathing cycles, which might have different breathing amplitudes. Once scans are divided into phase bins, although corresponding to the same breathing phase, they might correspond to different amplitudes, and therefore diaphragm position in one of those bins might not correspond to the tumor position identified in the same bin. We believe that a study that uses real time position is more accurate than using 4D-CT data.

This paper only intended to evaluate correlation of tumor and diaphragm. Variability of the model correlation and slope throughout different sequences has been analyzed in several patients. A future study will evaluate whether the model built for one treatment fraction can predict motion during a different fraction or inter-fraction variation of the correlation is too large to use the model from one fraction for prediction in another fraction.

In summary, radiotherapy treatments based on surrogate imaging rely on the good correlation of the tumor motion with respect to the surrogate. We have seen in the present study that this correlation is good when using diaphragm as a surrogate, and that it can be improved by using more sophisticated correlation models. However, this correlation is patient specific, and it should be examined for each patient individually before it is used for deriving the tumor position.

**Acknowledgements**

The authors would like to thank Sonia Gupta for her help with the image processing.

**References**

Arslan S, Yilmaz A, Bayramgurler B, Uzman O, Nver E and Akkaya E 2002 CT-guided transthoracic fine needle aspiration of pulmonary lesions: accuracy and complications in 294 patients *Med Sci Monit* **8** CR493-7

Balter J M, Wright J N, Newell L J, Friemel B, Dimmer S, Cheng Y, Wong J, Vertatschitsch E and Mate T P 2005 Accuracy of a wireless localization system for radiotherapy *Int J Radiat Oncol Biol Phys* **61** 933-7

Berbeco R I, Mostafavi H, Sharp G C and Jiang S B 2005a Towards fluoroscopic respiratory gating for lung tumours without radiopaque markers *Phys Med Biol* **50** 4481-90

Berbeco R I, Nishioka S, Shirato H, Chen G T and Jiang S B 2005b Residual motion of lung tumours in gated radiotherapy with external respiratory surrogates *Phys Med Biol* **50** 3655-67

Berbeco R I, Nishioka S, Shirato H and Jiang S B 2006 Residual motion of lung tumors in end-of-inhale respiratory gated radiotherapy based on external surrogates *Med Phys* **33** 4149-56

Bruce E N 1996 Temporal variations in the pattern of breathing *J Appl Physiol* **80** 1079-87

Chen Q S, Weinhous M S, Deibel F C, Ciezki J P and Macklis R M 2001 Fluoroscopic study of tumor motion due to breathing: facilitating precise radiation therapy for lung cancer patients. *Med Phys* **28** 1850-6

Cui Y, Dy J G, Sharp G C, Alexander B and Jiang S B 2007 Multiple template-based fluoroscopic tracking of lung tumor mass without implanted fiducial markers *Phys Med Biol* **52** 6229-42

Fang L C, Komaki R, Allen P, Guerrero T, Mohan R and Cox J D 2006 Comparison of outcomes for patients with medically inoperable Stage I non-small-cell lung cancer treated with two-dimensional vs. three-dimensional radiotherapy *Int J Radiat Oncol Biol Phys* **66** 108-16

Geraghty P R, Kee S T, McFarlane G, Razavi M K, Sze D Y and Dake M D 2003 CT-guided transthoracic needle aspiration biopsy of pulmonary nodules: needle size and pneumothorax rate *Radiology* **229** 475-81

Hoisak J D, Sixel K E, Tirona R, Cheung P C and Pignol J P 2004 Correlation of lung tumor motion with external surrogate indicators of respiration *Int J Radiat Oncol Biol Phys* **60** 1298-306

Hoisak J D, Sixel K E, Tirona R, Cheung P C and Pignol J P 2006 Prediction of lung tumour position based on spirometry and on abdominal displacement: accuracy and reproducibility *Radiother Oncol* **78** 339-46

Jain A K, Duin R P W and Jianchang M 2000 Statistical pattern recognition: a review *Pattern Analysis and Machine Intelligence, IEEE Transactions on* **22** 4-37

Jiang S B 2006a Radiotherapy of mobile tumors *Semin Radiat Oncol* **16** 239-48

Jiang S B 2006b Technical aspects of image-guided respiration-gated radiation therapy *Med Dosim* **31** 141-51

Kanoulas E, Aslam J A, Sharp G C, Berbeco R I, Nishioka S, Shirato H and Jiang S B 2007 Derivation of the tumor position from external respiratory surrogates with periodical updating of the internal/external correlation *Phys Med Biol* **52** 5443-56

Keall P J, Kini V R, Vedam S S and Mohan R 2001 Motion adaptive x-ray therapy: a feasibility study. *Phys Med Biol* **46** 1-10

Keall P J, Mageras G S, Balter J M, Emery R S, Forster K M, Jiang S B, Kapatoes J M, Low D A, Murphy M J, Murray B R, Ramsey C R, Van Herk M B, Vedam S S, Wong J W and Yorke E 2006 The management of respiratory motion in radiation oncology report of AAPM Task Group 76 *Med Phys* **33** 3874-900

Lin T, Cerviño L I, Tang X, Vasconcelos N and Jiang S B 2009 Fluoroscopic tumor tracking for image-guided lung cancer radiotherapy *Physics in Medicine and Biology* **54** 981-92

Nelson C, Starkschall G, Balter P, Morice R C, Stevens C W and Chang J Y 2007 Assessment of lung tumor motion and setup uncertainties using implanted fiducials *Int J Radiat Oncol Biol Phys* **67** 915-23

Pudil P, Novovičová J and Kittler J 1994 Floating search methods in feature selection *Pattern Recognition Letters* **15** 1119-25

Ruan D, Fessler J A, Balter J M, Berbeco R I, Nishioka S and Shirato H 2008 Inference of hysteretic respiratory tumor motion from external surrogates: a state augmentation approach *Phys Med Biol* **53** 2923-36

Seiler P G, Blattmann H, Kirsch S, Muench R K and Schilling C 2000 A novel tracking technique for the continuous precise measurement of tumour positions in conformal radiotherapy *Phys Med Biol* **45** N103-10.

Sharp G C, Jiang S B, Shimizu S and Shirato H 2004 Tracking errors in a prototype real-time tumour tracking system *Phys Med Biol* **49** 5347-56

Tang X, Sharp G C and Jiang S B 2007 Fluoroscopic tracking of multiple implanted fiducial markers using multiple object tracking *Phys Med Biol* **52** 4081-98

Tsunashima Y, Sakae T, Shioyama Y, Kagei K, Terunuma T, Nohtomi A and Akine Y 2004 Correlation between the respiratory waveform measured using a respiratory sensor and 3D tumor motion in gated radiotherapy *Int J Radiat Oncol Biol Phys* **60** 951-8 van der Weide L, van Sornsen de Koste J R, Lagerwaard F J, Vincent A, van Triest B, Slotman B J and Senan S 2008 Analysis of carina position as surrogate marker for delivering phase-gated radiotherapy *Int J Radiat Oncol Biol Phys* **71** 1111-7

Vedam S S, Kini V R, Keall P J, Ramakrishnan V, Mostafavi H and Mohan R 2003 Quantifying the predictability of diaphragm motion during respiration with a noninvasive external marker *Med Phys* **30** 505-13.

Wu H, Zhao Q, Berbeco R I, Nishioka S, Shirato H and Jiang S B 2008 Gating based on internal/external signals with dynamic correlation updates *Phys Med Biol* **53** 7137-50

Xu Q, Hamilton R J, Schowengerdt R A, Alexander B and Jiang S B 2008 Lung Tumor Tracking in Fluoroscopic Video Based on Optical Flow *Medical Physics* **35** 5351-9

Xu Q, Hamilton R J, Schowengerdt R A and Jiang S B 2007 A deformable lung tumor tracking method in fluoroscopic video using active shape models: a feasibility study *Phys Med Biol* **52** 5277-93

Zhang Q, Pevsner A, Hertanto A, Hu Y C, Rosenzweig K E, Ling C C and Mageras G S 2007 A patient-specific respiratory model of anatomical motion for radiation treatment planning *Med Phys* **34** 4772-81